\documentclass[10pt]{article}
\usepackage[top=0.85in,left=0.75in,footskip=0.75in,marginparwidth=2in]{geometry}

\usepackage[utf8]{inputenc}
\usepackage{booktabs}
\usepackage{cite}
\usepackage{longtable}
\usepackage{afterpage}
\usepackage{subcaption}
\usepackage{booktabs}
\usepackage{hhline}
\usepackage{rotating}
\usepackage{placeins}
\usepackage{comment}
\usepackage{nameref,hyperref}

\newtheorem{instruction}{Instruction}

\usepackage[right]{lineno}

\usepackage{microtype}
\DisableLigatures[f]{encoding = *, family = * }

\usepackage{changepage}

\usepackage[aboveskip=1pt,labelfont=bf,labelsep=period,singlelinecheck=off]{caption}

\makeatletter
\renewcommand{\@biblabel}[1]{\quad#1.}
\makeatother

\usepackage{lastpage,fancyhdr,graphicx}
\usepackage{epstopdf}
\fancyheadoffset[L]{2.25in}
\fancyfootoffset[L]{2.25in}

\usepackage{color}

\definecolor{Gray}{gray}{.25}

\usepackage{graphicx}
\usepackage{adjustbox}

\usepackage{sidecap}
\usepackage{csquotes}
\MakeOuterQuote{"}

\usepackage{wrapfig}
\usepackage[pscoord]{eso-pic}
\usepackage[fulladjust]{marginnote}
\reversemarginpar

\begin{document}
\vspace*{0.35in}

\begin{flushleft}
{\Huge
\textbf\newline{Cognitive network science reveals bias in GPT-3, ChatGPT, and GPT-4 mirroring math anxiety in high-school students}
}
\newline
\\
\textbf{Katherine Abramski$^{1,x}$, Salvatore Citraro$^{2,x}$, Luigi Lombardi$^{3}$, Giulio Rossetti$^{2,+}$, Massimo Stella$^{3,+}$}
\newline
\\
1. University of Pisa, Italy
\\
2. ISTI - CNR
\\
3. University of Trento, Italy
\\
\textbf{$^{x,+}: $These authors contributed equally.}
\\
\textbf{Corresponding author: mass.stella@unitn.it}
\bigskip

\end{flushleft}



{\small
\noindent Large language models are becoming increasingly integrated into our lives. Hence, it is important to understand the biases present in their outputs in order to avoid perpetuating harmful stereotypes, which originate in our own flawed ways of thinking. This challenge requires developing new benchmarks and methods for quantifying affective and semantic bias, keeping in mind that LLMs act as psycho-social mirrors that reflect the views and tendencies that are prevalent in society. One such tendency that has harmful negative effects is the global phenomenon of anxiety toward math and STEM subjects. Here, we investigate perceptions of math and STEM fields provided by cutting-edge language models, namely GPT-3, Chat-GPT, and GPT-4, by applying an approach from network science and cognitive psychology. Specifically, we use behavioral forma mentis networks (BFMNs) to understand how these LLMs frame math and STEM disciplines in relation to other concepts. We use data obtained by probing the three LLMs in a language generation task that has previously been applied to humans. Our findings indicate that LLMs have an overall negative perception of math and STEM fields, with math being perceived most negatively. We observe significant differences across the three LLMs. We observe that newer versions (i.e. GPT-4) produce richer, more complex perceptions as well as less negative perceptions compared to older versions and $N=159$ high-school students. These findings suggest that advances in the architecture of LLMs may lead to increasingly less biased models that could even perhaps someday aid in reducing harmful stereotypes in society rather than perpetuating them.
}
\\

\section{Introduction}

The introduction of large language models (LLMs) has taken the world by storm, and society's reaction has been anything but unanimous, ranging from humorous amusement to catastrophic fear. Among the most prominent LLMs are OpenAI's GPT-3, ChatGPT, and GPT-4 (ordered oldest to newest). GPT-3 and GPT-4 are powerful and flexible models that can be fine-tuned to perform a wide variety of natural language processing tasks, while ChatGPT is a variant of the other two specifically designed to perform well in conversational contexts. All three belong to the family of the generative pre-trained transformer (GPT) models \cite{openai2023gpt4} that are trained on massive amounts of textual data to learn patterns and relationships in text \cite{brown2020language,ESOUZA2023119124}. Their power and versatility for accomplishing a range of tasks with incredible human-like finesse have led to a boom in their popularity in society and among researchers across disciplines.

As LLMs secure their role in our lives as useful tools for everyday tasks such as composing emails, writing essays, debugging code, and answering questions, the need to understand the behavior and risks of these models is ever more important \cite{binz2023using, shiffrin2023probing, srivastava2022beyond, hagendorff2023machine}. There has been a spike in research dedicated to this topic, surrounded by a debate about the nature of the capabilities of LLMs \cite{mitchell2023debate}. Some researchers have suggested that the impressive performance of LLMs on difficult reasoning tasks is indicative of an early version of general artificial intelligence \cite{bubeck2023sparks}. Many others argue that LLMs exhibit nothing like true understanding because they lack a grasp of meaning \cite{bender2020climbing}, arguing that they perform well but for the wrong reasons \cite{mitchell2023debate}. In fact, much of the success of LLMs at human-like reasoning tasks can be attributed to spurious correlations rather than actual reasoning capabilities \cite{niven2019probing}.

Despite opposing views regarding the nature of intelligence exhibited by LLMs, a relatively undisputed topic is the issue of bias. Bias, in the context of LLMs, has recently been studied as the presence of misrepresentations and distortions of reality that result in favouring certain groups or ideas, perpetuating stereotypes, or making incorrect assumptions \cite{ferrara2023chatgpt}. While these biases can be influenced by many factors, they largely originate from biases in the massive text corpora on which the models are trained. This can be due to certain groups or ideas being underrepresented in the training data, or from implicit biases present in the training data itself. Thus, the output produced by LLMs inevitably reflects stereotypes and inequalities prevalent in society. This is problematic since exposure through interaction with LLMs could lead to perpetuating existing stereotypes and even the creation of new ones \cite{caliskan2017semantics, ferrara2023chatgpt}.

As LLMs become more integrated into our lives, it is ever more important to investigate the biases produced by them. This includes understanding our own human biases as well, since LLMs act as "psycho-social mirrors" \cite{sasson2023mirror} that reflect human features of personality as well as societal views and tendencies. Thus, it is important to investigate the individual cognitive sphere in conjunction with LLM behavior to understand how our individual and societal tendencies are diffused into the knowledge possessed by artificial intelligence agents. A very natural yet negative human phenomena is affective bias \cite{anoop2022towards}, the tendency to prioritize the processing of emotionally negative events compared to positive ones \cite{pulcu2017affective}. An example of affective bias is attributing negative attitudes to neutral concepts, such as the attribution of negative perceptions to the neutral concept \textit{math}. These types of biases and stereotypes are inherited by LLMs, adopting perceptions of neutral concepts that deviate significantly from neutrality as a result of our own biased perceptions. It should be the goal of researchers working on developing LLMs to understand such nuanced biases in humans to ensure that LLMs adopt neutral unbiased views of concepts or phenomena that have been historically stigmatized or misrepresented. In doing so, regular widespread interaction with LLMs might actually contribute to a reduction in the harmful biases held by humans.


In this work, we investigate biases produced by LLMs, specifically GPT-3, ChatGPT, and GPT-4, regarding their perception of academic disciplines, particularly math, science, and other STEM fields. 
In many societies, these disciplines have a reputation for being difficult \cite{foley2017math}. Math in particular, which is arguably the language of science, has been known to cause a great deal of anxiety in many people. This anxiety is a global phenomenon \cite{foley2017math, luttenberger2018spotlight}, and it is deeply rooted, beginning in childhood and persisting throughout adulthood. Unpleasant feelings about math may already begin to develop as early as first grade \cite{maloney2012math}. Children pick up on the anxieties of their teachers and parents \cite{ramirez2018teacher}, similar to how LLMs absorb biases from training data. Unfortunately, negative perceptions of math have become so commonplace that it is not unusual to hear people identify themselves as not "math people". While this kind of self-categorization may seem harmless, math anxiety can actually have severe individual and societal consequences \cite{ashcraft2002math, ashcraft2005math,foley2017math,daker2021first}. Math anxiety may cause individuals to avoid situations in which math is involved, ultimately having a negative impact on performance. This avoidance tendency may cause bright and capable students to avoid math-intensive classes, determining the course of their academic and professional career \cite{daker2021first}. This scales to the societal level. Math anxiety may deter a large portion of the workforce from pursuing careers in STEM, which are in high demand, and since math anxiety is more prevalent in females as a result of societal stereotypes \cite{hembree1990nature}, it may contribute greatly to the gender gap in STEM fields.

Just as children are likely to mirror the math anxiety expressed by their teachers or parents \cite{stella2022network}, LLMs are "psycho-social mirrors" \cite{mitchell2023debate,sasson2023mirror}, which reflect the tone of the language that we use to talk about math. Thus, we expect to find negative attitudes towards math in large language models. It is critical to investigate the nature of this bias in order to identify ways to overcome it as AI architectures become more advanced.

To accomplish this, we apply behavioral forma mentis networks (BFMNs) as a method of investigation. BFMNs are a type of cognitive network model that capture how concepts are perceived by individuals or groups by building a network of conceptual associations \cite{stella2019forma}.
This framework, which arises from cognitive psychology coupled with tools from network science, can also be applied to probe LLMs to reveal how they frame concepts related to math, science, and STEM. In this study, we investigate perceptions of these disciplines in three LLMs: GPT-3, ChatGPT, and GPT-4. A comparison of these models allows us to gain a temporal perspective about how these biases may evolve as subsequent versions of these LLMs are released.

The rest of the paper is organized as follows. In Section \ref{sec:rev} we provide a review of recent research dedicated to investigating bias in language models, discussing benchmarks and methods for conducting psychological investigations of LLMs.
In Section \ref{sec:meth}, we describe the framework of BFMNs, and we provide details about data collection, analysis and visualization.
In Section \ref{sec:res}, we summarize the results of our investigation of bias towards academic disciplines present in the output from GPT-3, ChatGPT, and GPT-4, and in Section \ref{sec:disc}  we discuss the implications of our findings.

\section{Review of Recent Literature}
\label{sec:rev}

Bias has been a significant obstacle to the distributed approach to semantic representation from early on. Since the introduction of word embeddings like word2vec \cite{mikolov2013distributed}, researchers have been aware that the advantageous operations provided by these models, such as using vector differences to represent semantic relations, are likely to express undesired biases. For example, sexist and racist word analogies such as \textit{"father" is to "doctor" as "mother" is to "nurse"} \cite{bolukbasi2016man} and \textit{black is to criminal as Caucasian is to police} \cite{manzini2019black} produced by word embeddings demonstrate how language contains biases that reflect adverse societal stereotypes. Unfortunately, these types of biases are present in tools that we use every day. For example, Google Translate has been found to overrepresent males when translating from gender-neutral languages to English, especially in male-dominated areas such as STEM fields, perpetuating existing gender imbalances \cite{prates2020assessing}.

Cutting-edge LLMs like GPT-3, ChatGPT, and GPT-4 are not immune to these types of dangers, and the facility of LLMs to simulate human-like language-related competencies, like ChatGPT's tremendous ability in question answering and storytelling, makes it necessary to investigate the behavior of LLMs. This has led to the development of new methods and benchmarks for investigating bias that shed light on the variety of demographic and cultural stereotypes and misrepresentations present in the output of language models \cite{nadeem2020stereoset, ferrara2023chatgpt}.

Gender, racial, and religious stereotypes are among the most widely investigated biases. These biases can be detected in several ways, often by prompting the language model to generate language and then evaluating the output in several ways. One approach involves using Association Tests \cite{greenwald1998measuring, caliskan2017semantics, kurita2019measuring, nadeem2020stereoset}, which may be done at different levels of discourse. For example, at the word level, the strength of the association of two words like \textit{sister} and \textit{science} can be measured \cite{caliskan2017semantics}, providing a simple and intuitive way to measure bias in word embeddings. At the sentence level, the model may be prompted to complete a sentence like \textit{girls tend to be more \rule{1cm}{0.10mm} than boys}, or to make assumptions following a given context like \textit{He is an Arab from the Middle East}. \cite{nadeem2020stereoset}.

Similar approaches have been applied to investigate different types of bias in various LLMs from BERT and RoBERTa to GPT-3 and ChatGPT. Persistent anti-muslim bias has been detected by probing GPT-3 in various ways including prompt completion, analogical reasoning, and story generation \cite{abid2021persistent}. Topic modeling and sentiment analysis techniques have been used to find gender stereotypes in narratives generated by GPT-3 \cite{lucy2021gender}. Sentiment scores and measurements of "regard" towards a demographic have been applied to assess stereotypes related to gender and sexual orientation in output produced by GPT-2 \cite{sheng2019woman}.

While some biases are easier to spot, others are more nuanced \cite{magee2021intersectional} and hidden deeply in the architecture of LLMs. Tools from cognitive psychology may be better suited for detecting the subtler dangers of language models where performance-based methods fall short \cite{srivastava2022beyond, hagendorff2023machine, binz2023using}. For example, one may ask whether a chatbot like ChatGPT can manifest dangerous psychological traits or personalities when asked if it agrees or disagrees with statements like \textit{I am not interested in other people's problems} or \textit{I hate being the center of attention} \cite{li2022gpt}. Such psychological investigations can measure the extent to which LLMs inherently manifest negative personalities and dark connotations like Machiavellianism and narcissism \cite{li2022gpt}. Such investigations are an example of the emerging field of "machine psychology" \cite{hagendorff2023machine}, which applies tools from cognitive psychology to investigate the behavior of machines as if they were human participants in psychological experiments. The goal of this new field is to investigate the emergent capabilities of language models where traditional NLP benchmarks are insufficient.

\section{Methods}
\label{sec:meth}

Given that our method of investigation can be applied to both humans and LLMs, our approach using behavioral forma mentis networks (BFMNs) can be considered a type of "machine psychology". Combining knowledge structure and affective patterns, forma mentis networks identify how concepts are associated and perceived by individuals or populations. Here we build BFMNs out of free associations data and valence estimates produced by OpenAI's large language models GPT-3, ChatGPT, and GPT-4.

BFMNs represent ways of thinking as a cognitive network of interconnected nodes/concepts. Connections/links represent conceptual similarities or relationships. In BFMNs, links indicate memory recall patterns between concepts, which in this case are obtained through a free association game. In this cognitive task, an individual is presented with a cue word and asked to generate immediate responses to it, "free" from any detailed correspondence (responses need not be synonyms with the cue word). These free associations represent memory recall patterns, which can be represented as a network. For example, reading \textit{math} may make one think of \textit{number}, so the link (\textit{math}, \textit{number}) is established. In continued free association tasks \cite{de2013better}, up to three responses to a given cue can be recorded. Responses are not linked to each other, instead, they are connected only to the cue word. This maximizes the explanatory power that cognitive networks have in terms of explaining variance across a variety of language-processing tasks related to human memory (see \cite{de2013better}). Importantly, BFMNs are feature-rich networks, in that their network structure is enriched by node-level features expressing the valence of each concept, i.e. how positively or negatively a given concept is perceived by an individual or group.

Rather than building BFMNs from responses provided by humans, as done in previous works \cite{stella2019forma,stella2020forma,stella2020education}, in this study, BFMNs are constructed out of responses from textual interactions with language models. The same methodology is applied for GPT-3, ChatGPT, and GPT-4. The resulting networks thus represent how each LLM associates and perceives key concepts related to math, science, and STEM fields based on their responses to the language generation task.

\subsection{Data collection: Free associations and valence norms}

As a language generation task, we implemented a continued free association game \cite{de2013better}, providing each of the three language models with the following prompt, substituting different cue words:

\begin{instruction}
Write a list of 3 words that come to your mind when you think of \textit{CUE\_WORD} and rate each word on a scale from 1 (very negative) to 5 (very positive) according to the sentiment the word inspires in you.
\label{def:instruction}
\end{instruction}

For each prompt, the language model responded by providing three textual responses coupled with three related numerical responses (valence scores) between 1 and 5. Punctuation and blank spaces were manually removed. In addition to valence scores corresponding to the responses, we also asked each language model to provide a single valence score (independently evaluated) from 1 to 5 for each of the cue words. The language model failed to comply with the instructions only 5\% of the time, producing repetitions of the cue word in the response. Those instances were discarded and did not count as repetitions.

In a similar study performed on high school students \cite{stella2019forma}, there were 159 participants, each providing around three responses to each cue word. Therefore in this study, for comparison purposes, we repeated the above instructions to obtain at least 159 responses for each cue word, matching the number of students who took part in the human study. For GPT-3, we selected the DaVinci model with a temperature of $T=0.7$, which is the default setting. Iterations were automated in Python through the API service provided by OpenAI. Therefore, we obtained three datasets, one for each of the language models tested, with sample sizes comparable to that of the human dataset from \cite{stella2019forma}. This enabled interesting comparisons between the recollection patterns of language models and high school students.

To investigate attitudes towards math, science, and STEM subjects, we tested ten different cue words, corresponding to the same ten key concepts tested in the study with high school students \cite{stella2019forma}, namely: \textit{math}, \textit{physics}, \textit{science}, \textit{teacher}, \textit{scientist}, \textit{school}, \textit{biology}, \textit{art}, \textit{chemistry} and \textit{STEM}. Therefore, the above instructions can be read by substituting \textit{CUE\_WORD} with any of these ten key concepts (throughout this paper we use the terms \textit{key concept} and \textit{cue word} interchangeably).

For each key concept and its associated responses, valence scores (1 through 5) were converted into valence labels (\textit{negative}, \textit{positive}, or \textit{neutral}) using the Kruskall-Wallis non-parametric test (See Section \ref{sec:wordval} for details). Thus, valence could be considered categorically rather than numerically.


\subsection{Network building and semantic frame reconstruction}

Behavioral forma mentis networks (BFMNs) were constructed such that nodes represented lexical items and edges indicated free associations between words. Following the first part of Instruction \ref{def:instruction}, we built BFMNs as cognitive networks which simulate human memory recall patterns by linking the cue words to their associative responses. Given the selected cue words and the sets of three responses, our goal was to retrieve a network structure mapping how concepts were connected in the recall process, facilitated by the above instructions (see also \cite{stella2019forma}).

First of all, associative responses were converted to lowercase letters and checked automatically for common spelling mistakes. The automatic spell checkers used here were the ones implemented in Wolfram's Mathematica 11.3 (manufactured by Wolfram Research, Champaign, US). Secondly, different word forms were stemmed to reduce the occurrence of multiple word variants that convey the same concept. For stemming words, we used the WordStem command as implemented in Mathematica 11.3.

In the literature about semantic networks, there exist several ways to connect cue words to their associative responses \cite{de2013better,luchini2023convergent,de2019small}. We chose to connect each cue word to all three of its responses since this method has been shown to provide more heterogeneity in associative responses \cite{de2019small} and has been used in previous works with forma mentis networks \cite{stella2019forma,stella2020education,stella2021mapping}. Moreover, this approach to network construction has been shown to improve the accuracy of many language-related prediction tasks (such as associative strength prediction) compared to other strategies, e.g., connecting the cue word to the first response only \cite{de2019small}. 
We also considered idiosyncratic associations, i.e., associations provided only one time, which are visually represented as narrower edges compared to non-idiosyncratic associations.

Using the valence labels for the key concepts and associated responses, we enriched the BFMNs, representing them as feature-rich cognitive networks \cite{citraro2023feature} in which information about the sentiment of associative responses can be used to describe the properties of the cue word \cite{stella2019forma}. As in previous works, we leveraged the notion of a node's neighborhood, consisting of the set of adjacent nodes to a target node: In this case, the neighborhoods of a cue word are the sets of all the associative responses generated by the participants (the language models or humans) responding to the same set of instructions. Inspired by the famous quote \textit{You shall know a word by the company it keeps} \cite{firth1957synopsis}, which is also the foundation of the distributional semantic hypothesis \cite{lenci2018distributional}, we can get a better understanding of the valence attributed to the cue word by considering the valences of its neighboring associates.

\subsection*{Statistical analysis of word valence}
\label{sec:wordval}

For all key concepts and associated responses, in order to convert numerical valence scores (1 through 5) into categorical valence labels (\textit{negative}, \textit{positive}, or \textit{neutral}) we used a non-parametric statistical test. For each LLM, all valence scores provided for all key concepts and responses were aggregated together. A Kruskall-Wallis test was used to assess whether the scores attributed to concept $w_i$ had a lower, compatible, or higher median valence compared to the entire distribution of valence scores. Non-parametric testing was used because the distribution of valence scores $\bigcup_{j}w_{j}$ was mostly skewed with a heavy left tail across all models (Pearson's skewness coefficient $s_s=3(mean_s-median_s)/\sigma =$ 1.39 for students' data and $s_r>1$ for each language model).
Given the relatively small sample size (fixed in order to make suitable comparisons between large language models and humans), and inspired by previous works \cite{stella2019forma}, we fixed a significance level $\alpha = 0.1$, motivated by the aim of detecting more deviations from neutrality despite the contained sample size. Therefore, valence labels were assigned as follows: \textit{negative} - lower median valence score than the rest of the sample;  \textit{positive} - higher median valence score than the rest of the sample; \textit{neutral} - same median valence as the rest of the sample.

\subsection*{Data visualization, emotional analysis, and network neighborhood measurements}

In our network visualizations, we focused on reproducing the neighborhood of a given target concept, i.e. the associates corresponding to \textit{math}. We rendered valence through colors: Positive words were rendered in cyan, negative words in red, and neutral words in black. Idiosyncratic links were rendered with narrower edges compared to associated responses provided more than once. To better highlight clusters of associates, we used a hierarchical edge-bundling layout for network visualization. Because of space issues and to avoid overlap between node labels, we also used a star-graph layout. Both visualizations provide insights into the network structure of associates surrounding a key concept.

In this manuscript, we also used visualizations inspired by the circumplex model of affect \cite{posner2005circumplex}, which maps individual concepts as points in 2D dimensional space with valence and arousal. According to semantic frame theory \cite{fillmore2001frame} and distributional semantics in psycholinguistics \cite{malandrakis2013distributional}, each network neighborhood represents a semantic frame indicating ways in which a given concept is associated with others. Hence, understanding the distributions of valence and arousal scores attributed to associates in a given neighborhood provides crucial insights to better understanding how key concepts are perceived by a LLM or by a group of individuals \cite{poquet2020reviewing,stella2019forma}. For instance, in order to better understand the emotional content of the BFMN neighborhood surrounding \textit{math}, we can plot the 2D density plot for valence-arousal scores attributed to all words in the neighborhood, and then observe where associate words tend to cluster within the circumplex. We based these investigations on valence-arousal scores obtained by the National Research Canada Valence-Arousal-Dominance lexicon \cite{mohammad2018obtaining}.

Last but not least, we compared network neighborhoods, also called semantic frames, across large language models and humans. We measured the following aspects of a frame for each key concept $K$ across LLMs and high school students: (1) semantic frame size, i.e. the number of unique associates in the semantic frame; (2) estimated valence, i.e. the arithmetic mean of the valence scores attributed to $K$; (3) estimated frame valence, i.e. the mode of the valence labels attributed to the associates of $K$; (4) the fractions of positive/neutral/negative words present in the frame; (5) the fraction of non-emotional words present in the frame, i.e. the fraction of words that did not elicit any emotion (according to an emotion-word associative thesaurus \cite{mohammad2013crowdsourcing}) and could thus be considered as neutral domain-knowledge or technical associates to a key concept; (6) the fraction of positive/negative/neutral non-emotional words present in the frame.

\section{Results}
\label{sec:res}

This section outlines the key results we achieved in our interactions with large language models. To begin, we focus on the results from GPT-3 and ChatGPT. We start with an overall analysis of the valence patterns corresponding to each key concept for both LLMs. We then continue with a detailed analysis of the semantic frames surrounding each key concept, including an investigation of the content and valence of the associates, adopting an approach that uses a circumplex model of affect. Finally, we compare the results from GPT-3 and ChatGPT with results from GPT-4 to gain a better understanding of how LLMs are evolving as subsequent versions are released.

\subsection{Semantic frames of STEM concepts produced by GPT-3 and ChatGPT}

\begin{figure}[h!]
    \centering
    \includegraphics[width=14cm]{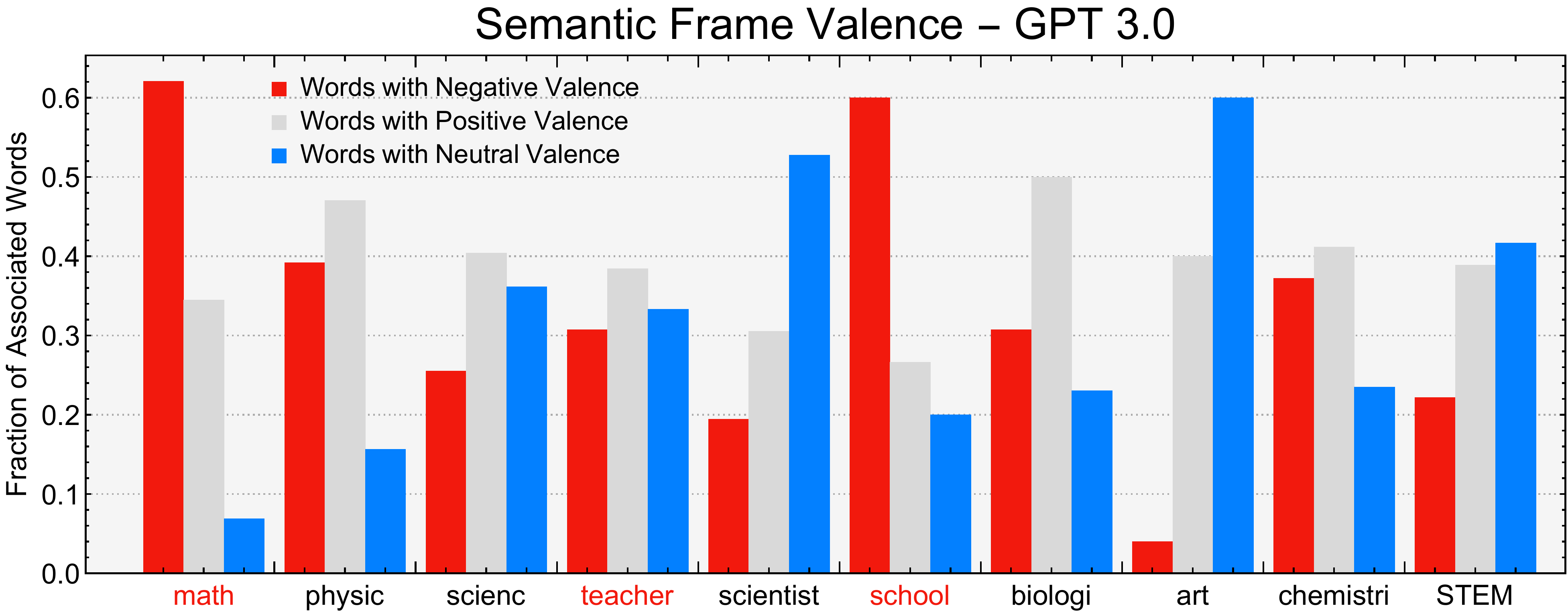}
    \includegraphics[width=14cm]{Figures/AurasCharts_ChatGPT.pdf}
    \caption{Fractions of positive, negative and neutral associated words populating the semantic frames of each cue word in GPT-3 (top) and ChatGPT (bottom).}
    \label{fig:charts}
\end{figure}

As discussed in the previous section, the LLMs were prompted to assign valence scores to all cue words and associated responses, and those valence scores were then converted to valence labels:   \textit{negative}, \textit{positive}, or \textit{neutral}. Figure \ref{fig:charts} reports the fraction of negative (red), positive (cyan), and neutral (gray) associated responses comprising the semantic frames of all ten cue words provided to ChatGPT and GPT-3. Cue words are reported at the bottom of each bar chart and colored according to the valence scores produced by each LLM. Notice that this coloring is independent of the valence polarity of the cue word's associates. The most frequent valence label in a semantic frame represents a connotation, also called a valence aura in \cite{stella2019forma}, not to be confused with a valence label. A valence label depends only on the scores attributed to that specific concept, while a valence aura/connotation depends on the valence labels of all the associates of that concept. Hence, whereas valence labels are applied only to individual concepts in isolation, valence connotations include information about network structure and thus constitute additional information about how a concept was perceived (see Methods). 

In terms of valence labels, as reported in Figure \ref{fig:charts} (top), GPT-3 did not identify any cue words as positive. The concepts \textit{math}, \textit{teacher}, and \textit{school} were all identified as negative concepts. Furthermore, \textit{math} and \textit{school}  were surrounded predominantly by other negative concepts, i.e. they had negative semantic frames \cite{stella2019forma}. The semantic frames provided by GPT-3 for \textit{physics}, \textit{chemistry}, and \textit{teacher} were highly polarized in terms of containing similar proportions of positive and negative associates, while \textit{art} and \textit{scientist} were associated mostly with neutral jargon (68$\%$ and 52$\%$, respectively). These patterns indicate a non-negligible amount of negative associations provided for \textit{math} and other academic concepts.

As reported in Figure \ref{fig:charts} (bottom), ChatGPT identified four of the ten cue words as positive concepts, and it provided considerably fewer negative semantic frames compared to GPT-3 overall. While both LLMs perceived \textit{math} negatively, the semantic frame of \textit{math} produced by ChatGPT contained 10\% fewer negative associates compared to that produced by GPT-3. Valence polarization was present in the semantic frames of \textit{physics}, \textit{science}, and \textit{chemistry} for ChatGPT. Notably, the number of negative associates corresponding to the concept \textit{STEM} provided by ChatGPT were nearly half those provided by GPT-3 ($24\%$ vs $15\%$).

The above findings provide evidence that both GPT-3 and ChatGPT not only perceived \textit{math} as a negative concept but also framed it negatively. There is also evidence of polarized semantic frames surrounding several concepts. Our findings warrant further investigation, so we proceed by investigating the semantic content of associations, aiming to understand how they would be emotionally interpreted by humans.

\subsection*{LLMs perceive \textit{math} much more negatively than \textit{science}}

\begin{figure}[h!]
    \centering
    \includegraphics[width=14cm]{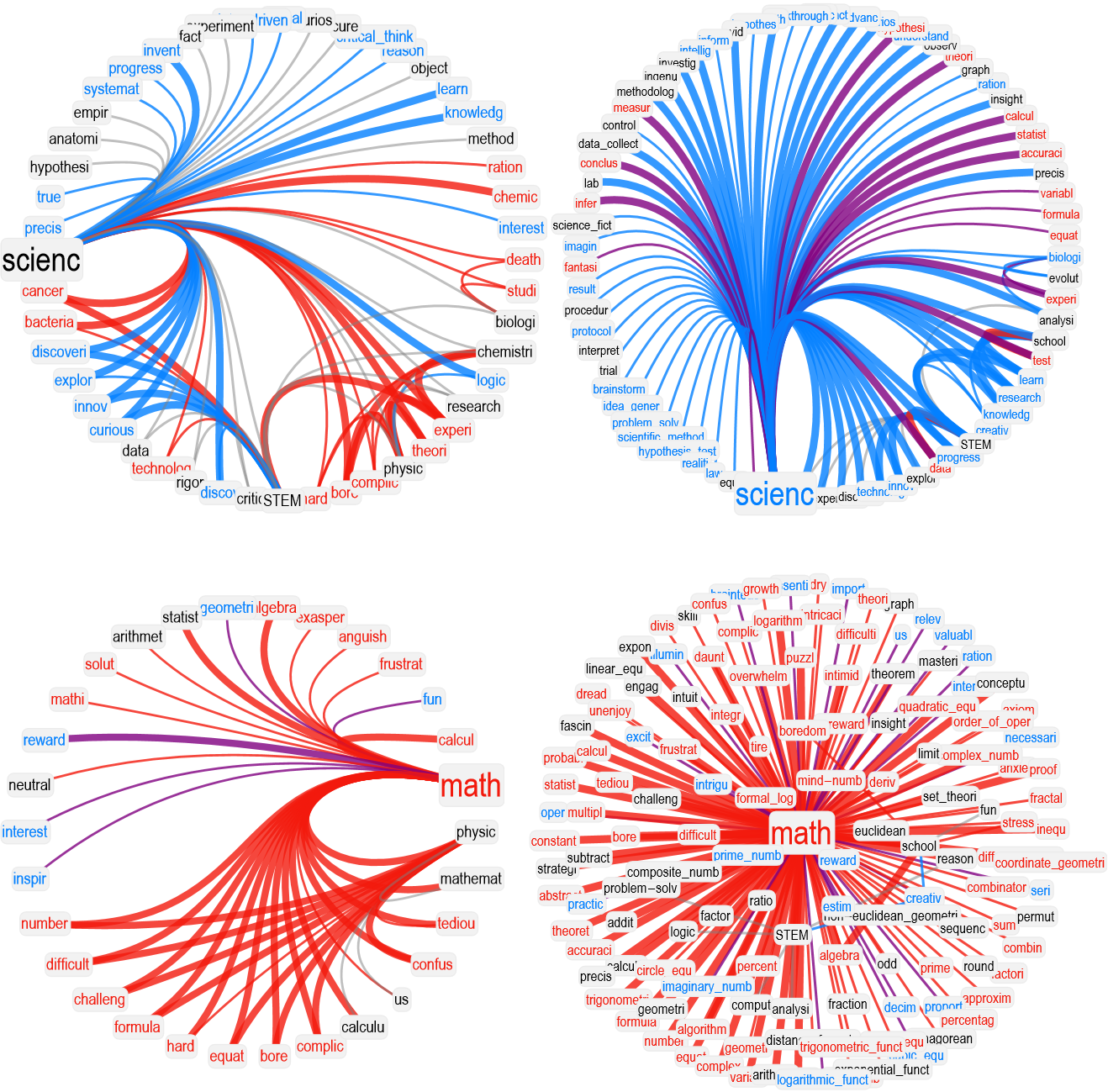}
    \caption{Semantic frames for \textit{science} (top) and \textit{math} (bottom) as produced by GPT-3 (left) and ChatGPT (right). Words in red (cyan) were rated as negative (positive). Words in black were rated as neutral. The cue word is displayed in a larger font size. Links between two negative terms are shown in red, while links between two positive terms are shown in cyan. Links between positive and negative words are shown in purple, indicating conflicting associations.}
    \label{fig:sciencemath}
\end{figure}

As discussed earlier, the words in the semantic frame of a key concept provide important contextual information about how that key concept is perceived. Figure \ref{fig:sciencemath} visualizes the semantic frames for \textit{science} (top) and \textit{math} (bottom) as produced by GPT-3 (left) and ChatGPT (right). 

GPT-3 framed \textit{science} mostly in neutral (\textit{data}, \textit{hypothesis}, \textit{method}) and positive (\textit{curious}, \textit{discover}, \textit{knowledge}) terms, although there is a non-negligible fraction of negative associates. Some of these negative associates relate to objects of investigation in science (e.g. \textit{bacteria}, \textit{chemicals}) while another cluster of negative associates describes science as \textit{hard}, \textit{boring} and \textit{complicated}. Noticeably, \textit{physics} also appears in this cluster.

ChatGPT framed \textit{science} in noticeably more positive terms. It is also easy to see that ChatGPT provided a richer, larger set of associates compared to GPT-3, which might depend on the more advanced level of sophistication achieved by ChatGPT compared to its predecessor. While the semantic frame of \textit{science} provided by ChatGPT is overwhelmingly positive, there are some negative associations that mostly relate to theoretical and mathematical aspects of science. In contrast to GPT-3, the terms \textit{complicated} and \textit{boring} were not present in the semantic frame produced by ChatGPT.

The concept \textit{math} was perceived and framed overwhelmingly negatively by both LLMs. As with \textit{science}, ChatGPT produced almost twice as many associates for \textit{math} compared to GPT-3, another indication of the more advanced competencies exhibited by ChatGPT.

GPT-3 framed \textit{math} as a \textit{boring}, \textit{difficult}, \textit{tedious}, \textit{frustrating} and \textit{exasperating} concept. Theoretical tools used in math, like \textit{equation} and \textit{formula}, were also perceived negatively. Such overwhelmingly negative perceptions were echoed by the associates provided by ChatGPT, which identified similar negative aspects of math, describing it as \textit{stressful}, \textit{complicated}, \textit{overwhelming}, and \textit{dreadful}. 

Compared to GPT-3, ChatGPT provided a considerably larger amount of associates related to domain knowledge for \textit{math}, reflecting a more advanced knowledge of mathematical tools, like \textit{exponentials}, \textit{fractions}, \textit{trigonometrics}, \textit{percentages}, and \textit{equations}, among others. Most associates from domain knowledge were perceived negatively, bolstering the overall negative connotation attributed by ChatGPT to \textit{math} in its semantic frame. 

The mixture of negative descriptive associations and negatively perceived domain knowledge terms provided by ChatGPT strongly echoes the negative semantic frame of \textit{math} provided by high school students identified in the previous work \cite{stella2019forma}. Our findings here provide strong evidence that vanilla ChatGPT provides a rather complex but strongly negative perception of math, which is consistent with some negative perceptions possessed by some student populations.

\begin{figure}
    \centering
    \includegraphics[width=12cm]{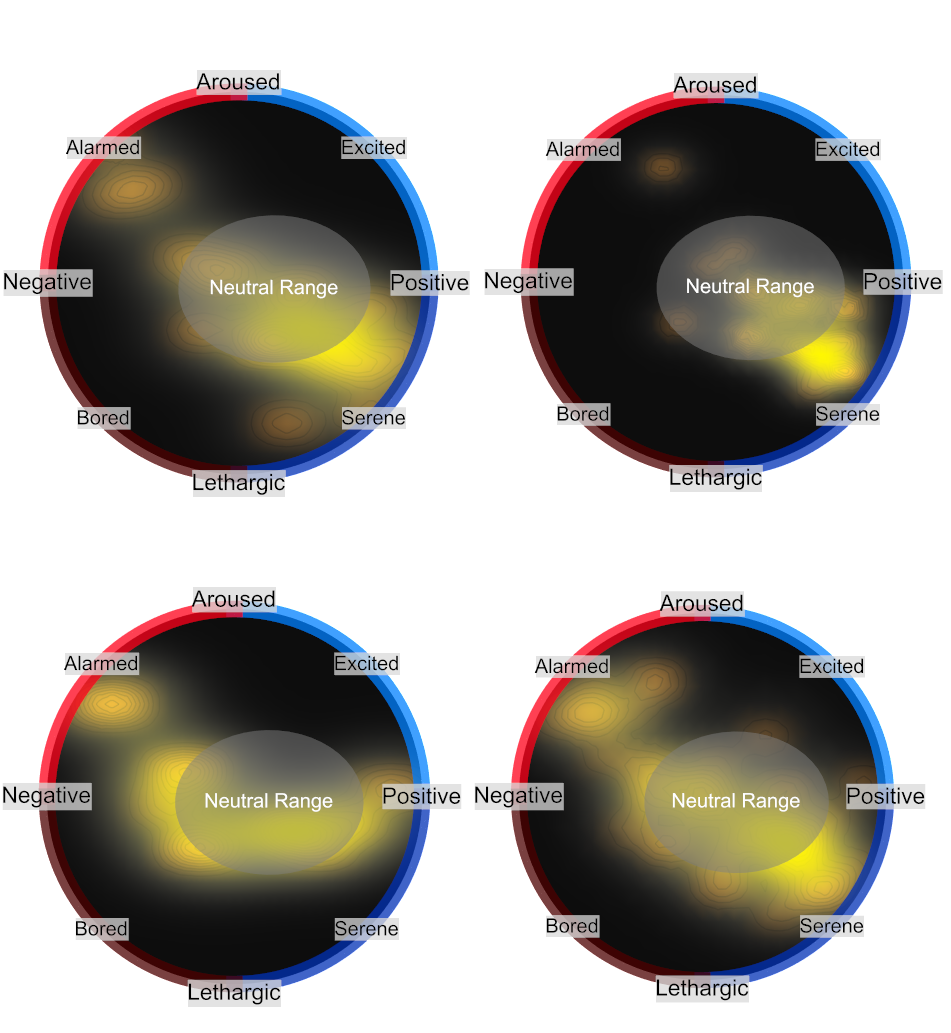}
    \caption{Circumplex model for the semantic frames of \textit{science} (top) and \textit{math} (bottom) as reproduced by GPT-3 (left) and ChatGPT (right).}
    \label{fig:sciencemathcirc}
\end{figure}

The above results depend on the valence scores provided by LLMs. To further assess the presence and extent of negative emotions in semantic frames, we also used external scores, namely, arousal scores based on human judgment (from \cite{mohammad2018obtaining}, see Methods) for assessing the emotional connotation of every associate. Figure \ref{fig:sciencemathcirc} reports the 2D distributions of valence and arousal scores for all concepts in the semantic frames of \textit{science} and \textit{math} as produced by GPT-3 and ChatGPT. A more intense yellow color indicates a concentration of associations within the same region of the circumplex model of affect \cite{russell1980circumplex}, where the dimensions of valence (x-axis) and arousal (y-axis) map different emotional states. Notice that these models identify how concepts in semantic frames would be emotionally perceived by humans, thus providing a different perspective compared to the valence scores provided by language models discussed so far.

The associates in the semantic frame of \textit{science} produced by both ChatGPT and GPT-3, Fig. \ref{fig:sciencemathcirc} (top right) and (top left), respectively,  are concentrated mostly in the lower right quadrant, which corresponds to emotions of serenity and tranquillity, i.e. positive valence and low arousal. This indicates that both the circumplex model and the forma mentis neighborhood portrayed \textit{science} as a concept inspiring calm. More negative associations persisted in GPT-3 for \textit{science}, as indicated by a cluster of concepts in the upper left quadrant, corresponding to anxiety and alertness, i.e. negative valence and higher arousal. This pattern, which is absent in ChatGPT, corresponds with the negative associations outlined in the above semantic frame analysis.

The distribution of concepts in the circumplex of affect is considerably different for \textit{math}. In Fig. \ref{fig:sciencemathcirc} (bottom left), GPT-3 features only a few concepts with positive valence scores and negligible arousal, falling outside the neutral range, a configuration considerably different from the one corresponding to \textit{science}. Several concepts fall in the upper-left quadrant, confirming the anxious perception of \textit{math} provided by vanilla GPT-3 which was discussed in the semantic frame analysis above.

ChatGPT framed \textit{math} in a considerably more emotionally polarized way (See Fig. \ref{fig:sciencemathcirc}, bottom right), compared to GPT-3. The associations provided by ChatGPT are distributed along a direction that spans the lower-right and the upper-left quadrants, combining emotions of serenity and tranquillity with alertness, anxiety, and alarm. This emotional polarization further underlines the complexity of ChatGPT, which is a language model capable of providing semantically richer and more emotionally polarized semantic frames for \textit{math} compared to GPT-3. Furthermore, both GPT-3 and ChatGPT frame \textit{math} in more negative terms compared to \textit{science}.

\subsection*{GPT-3 perceives \textit{school}  and \textit{teachers} much more negatively than ChatGPT}

\begin{figure}[h!]
    \centering
    \includegraphics[width=14cm]{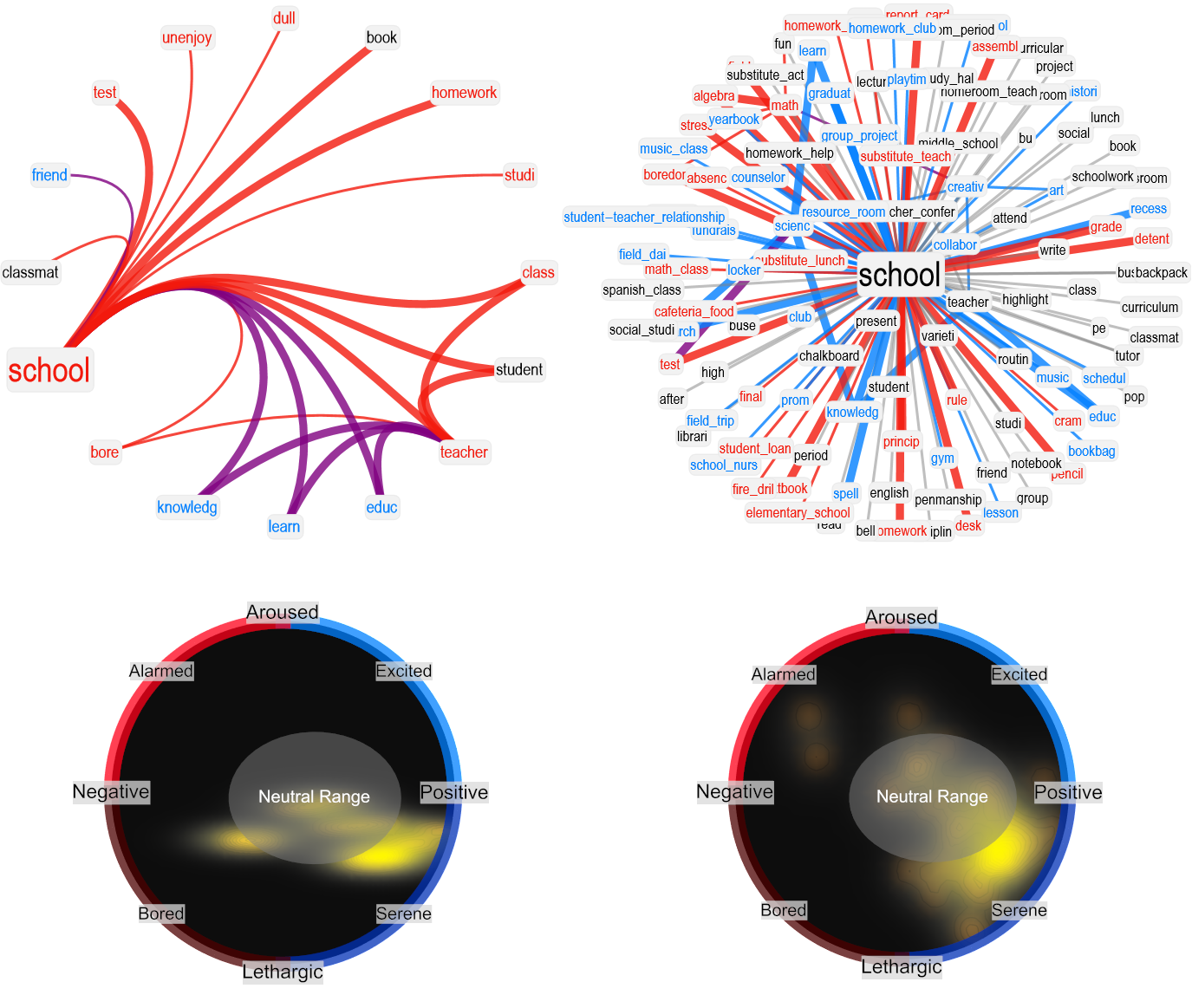}
    \caption{Semantic frames (top) and circumplex models (bottom) for \textit{school} as reproduced by GPT-3 (left) and ChatGPT (right). color patterns are the same as in Figure 2.}
    \label{fig:schoo}
\end{figure}

Figure \ref{fig:schoo} reports the semantic frames for \textit{school} for GPT-3 (top left) and ChatGPT (top right), together with their circumplex models (bottom). 

GPT-3 perceived \textit{school} as a negative concept and framed it with mostly negative and positive concepts. The language model associated school with positive jargon about learning (e.g. \textit{learn}, \textit{knowledge}, \textit{education}) but also with negative jargon about tests, boredom, and dullness (\textit{dull}, \textit{unenjoyable}, \textit{bored}). This dichotomy was confirmed also by the circumplex model, where most concepts fell in the lower-right quadrant (expressing serenity) and in the lower-left quadrant (expressing boredom). GPT-3 thus framed \textit{school} as a partly boring, partly serene concept, crucial for learning.

ChatGPT provided a much richer semantic frame for \textit{school} compared to GPT-3, mixing both positive (\textit{student-teacher relationship}, \textit{knowledge}, \textit{education}) and negative associates (\textit{cafeteria food}, \textit{detention}, \textit{algebra}, \textit{lunch}). Interestingly, the language model associated \textit{school} with \textit{algebra} and perceived the latter as a negative concept. Negative perceptions of \textit{algebra} represent one of several indicators of math anxiety, as captured by the psychometric scale detecting math anxiety developed by Hunt and colleagues \cite{hunt2011development}. Interestingly, the circumplex model for the semantic frame of \textit{school} does not reflect strong negative patterns, rather, most concepts in the model concentrate in the lower-right quadrant, corresponding to emotions of serenity and tranquillity. This dichotomy indicates that most of the negative associations reported in the semantic frame are due to the specific perceptions produced by ChatGPT and cannot be reflected or reproduced by how a large population of humans would perceive those same concepts. For example, ChatGPT might perceive \textit{algebra} as a negative concept while the NRC lexicon does not attribute negative valence scores to \textit{algebra}. Observing this difference further underlines the power of behavioral forma mentis networks in terms of adapting to the specific perceptions portrayed by specific individuals or groups.

\begin{figure}[h!]
    \centering
    \includegraphics[width=14cm]{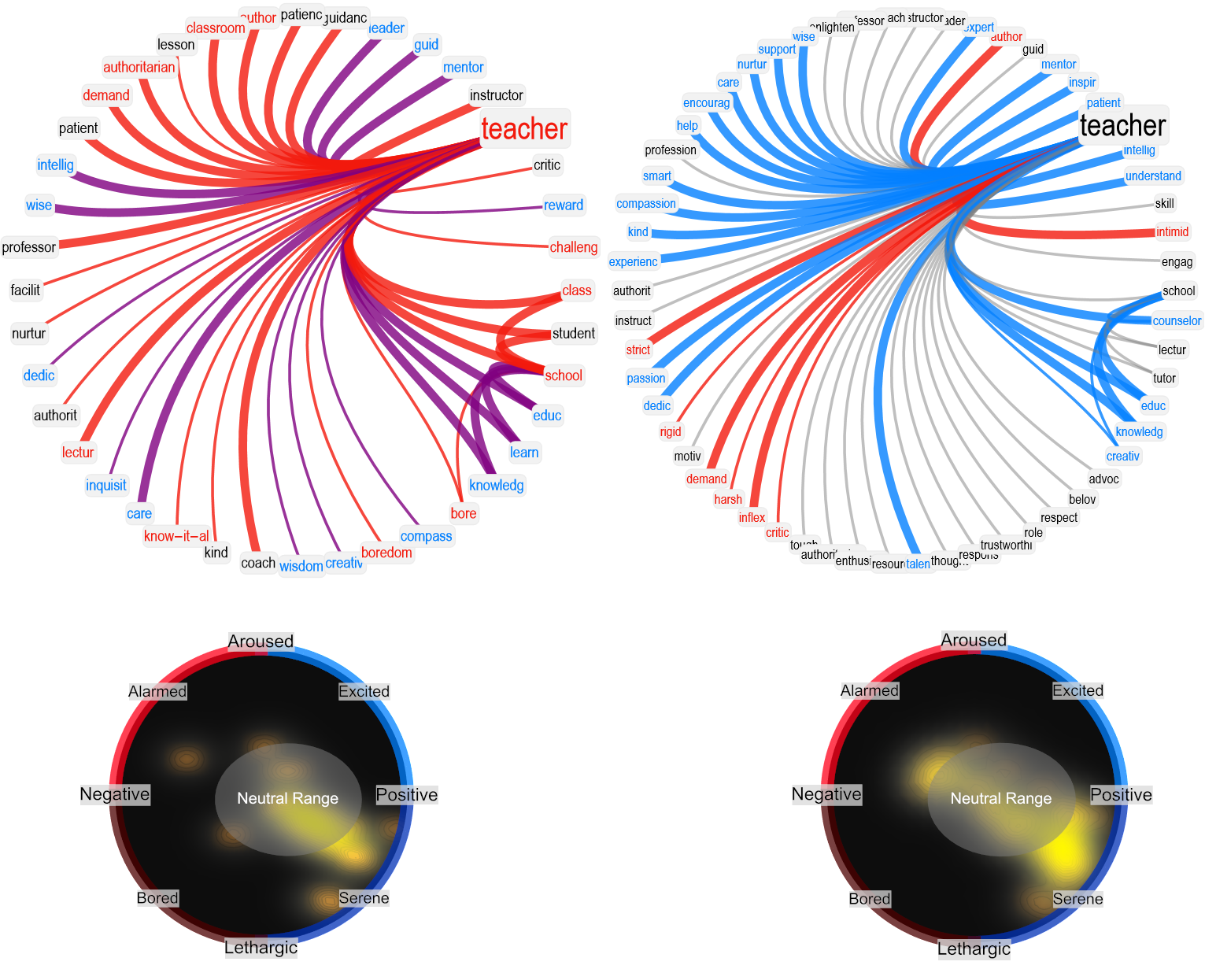}
    \caption{Semantic frames (top) and circumplex models (bottom) for \textit{teacher} as reproduced by GPT-3 (left) and ChatGPT (right). color patterns are the same as in Figure 2.}
    \label{fig:teac}
\end{figure}

Figure \ref{fig:teac} reports the semantic frames for \textit{teacher} for GPT-3 (top left) and ChatGPT (top right), together with their circumplex models (bottom). The semantic frames for \textit{teacher} differ significantly between GPT-3 and ChatGPT. GPT-3 identified "teacher" with a negative valence and surrounds it mostly with other negative concepts, e.g. \textit{authoritarian}, \textit{demanding}, \textit{know-it-all}, \textit{boredom}. These negative associations co-exist with positive ones, mentioning aspects related to knowledge transmission (e.g. \textit{education}, \textit{wisdom}, \textit{wise}) and mentoring (e.g. \textit{dedicate}, \textit{caring}, \textit{mentor}). This emotional polarity is also confirmed by the circumplex model.

Compared to GPT-3, ChatGPT put a stronger focus on the positive aspects of \textit{teacher}, underlining their ability to \textit{nurture}, \textit{encourage}, \textit{care} and \textit{support} in their roles. These aspects provide a perception of teachers as leaders and mentors, such that positive aspects dominate the entire semantic frame. However, negative perceptions are still present, i.e. associations between \textit{teacher} and \textit{strict}, \textit{demand}, \textit{harsh}, and \textit{criticize}. This dualistic positive/negative perception of teachers is also confirmed by the circumplex model, where concepts concentrate in the lower-right and upper-left quadrants, corresponding to emotions of calmness and alertness, respectively.

\subsection*{Other perceptions: \textit{physics}, \textit{chemistry}, \textit{STEM}, \textit{art} and \textit{scientist}}

\begin{figure}
    \centering
    \includegraphics[width=14cm]{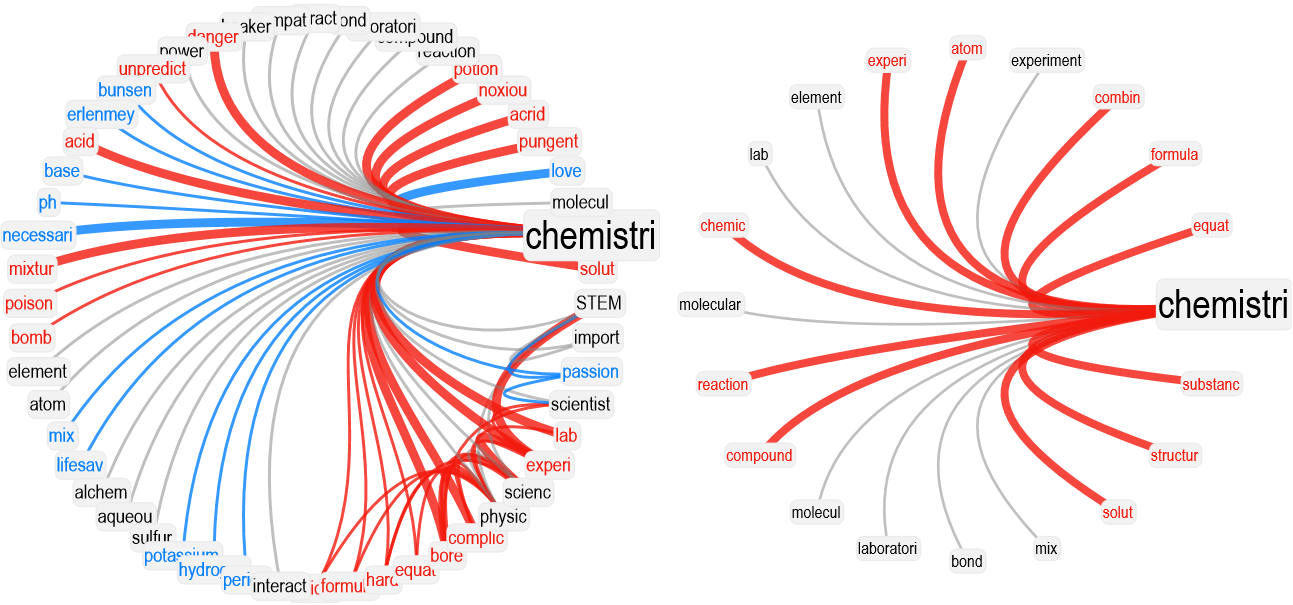}
    \includegraphics[width=14cm]{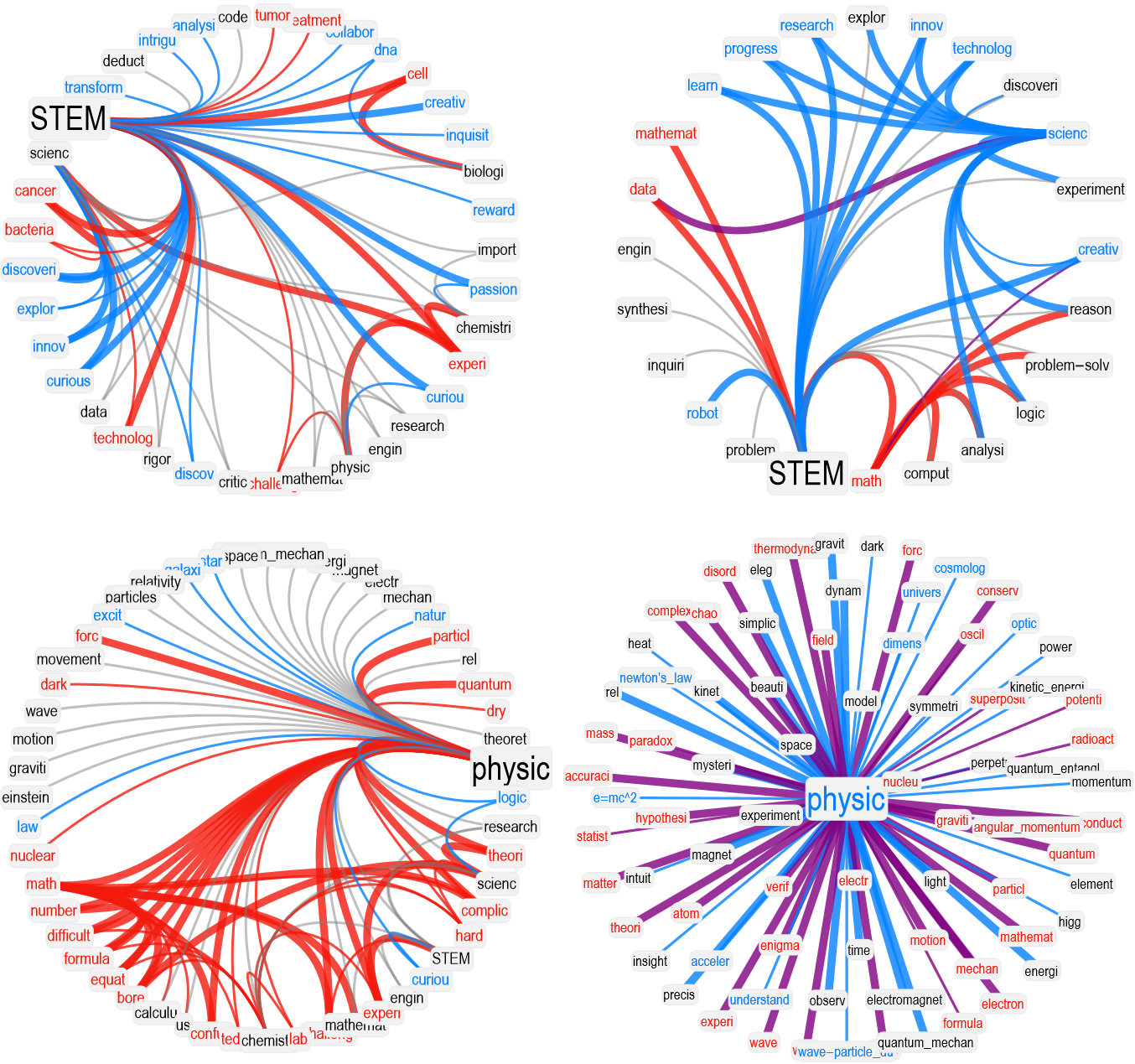}
    \caption{Semantic frames for \textit{chemistry} (top), \textit{STEM} (middle) and \textit{physics} (bottom) as reproduced by GPT-3 (left) and ChatGPT (right). color patterns are the same as in Figure 2.}
    \label{fig:stemphys}
\end{figure}

Figure \ref{fig:stemphys} portrays the semantic frames for \textit{chemistry} (top), \textit{STEM} (middle) and \textit{physics} (bottom).

\begin{figure}
    \centering
    \includegraphics[width=14cm]{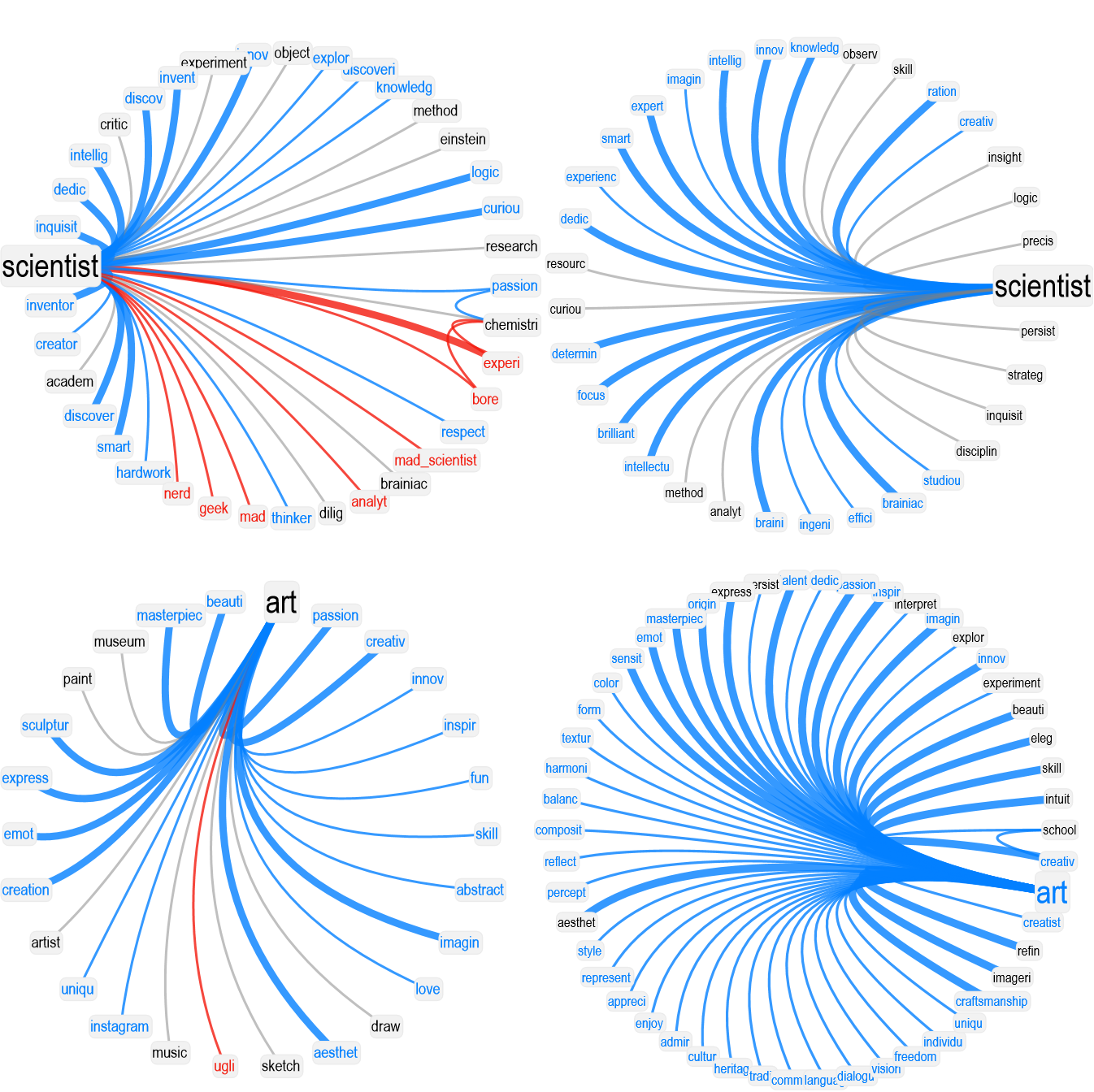}
    \caption{Semantic frames for \textit{scientist} (top) and \textit{art} (bottom) as reproduced by GPT-3 (left) and ChatGPT (right). color patterns are the same as in Figure 2.}
    \label{fig:scientistart}
\end{figure}

GPT-3 identified all three concepts as neutral but surrounds \textit{physics} with a mostly negative frame whereas semantic frames for both \textit{chemistry} and \textit{STEM} are polarized. Semantic network analysis reveals that the negative associates of \textit{chemistry} mostly relate to aspects of reagents and acids (\textit{poison}, \textit{pungent}, \textit{acrid}). Similarly, the negative associates of \textit{STEM} were mostly related to aspects in the health sciences where STEM can bring substantial improvements to well-being. In this way, the negativity found in the semantic frames for \textit{chemistry} and \textit{STEM} can be explained in terms of negative elements or challenges studied in these disciplines. This pattern is strikingly different from the negative associates found in the semantic frame of \textit{physics}, which mentions the concepts \textit{hard}, \textit{difficult} and \textit{complicated}. These associates are not elements studied in physics, but rather negative perceptions, which were found also in the semantic frames of \textit{math}.

ChatGPT associated \textit{chemistry} with general-level experimental and theoretical elements, mostly perceived as neutral or negative. This indicates another difference in how the two language models portray the same concept. Interestingly, ChatGPT produced a positive semantic frame for \textit{STEM}, establishing links with concepts like \textit{technology}, \textit{innovation}, and \textit{progress}. This finding is analogous to high school students holding negative perceptions of math and physics while holding positive attitudes towards science \cite{stella2019forma,stella2020forma,stella2020education}. ChatGPT framed \textit{STEM} in positive terms but also linked it with negatively perceived terms like \textit{math} and \textit{mathematical}. This dichotomy suggests that ChatGPT reflects distorted perceptions in which math is perceived negatively but still associated with positive perceptions of STEM and research.

Notably, ChatGPT perceived \textit{physics} as a positive entity, framed within a highly polarized semantic frame, rich with positive and negative concepts. This is in contrast with ChatGPT's negative perception of \textit{math}. This pattern differs from what was found in previous studies \cite{stella2019forma,stella2020forma,stella2020education}, where high school students framed both \textit{math} and \textit{physics} in negative terms.

The overwhelmingly positive semantic frames of \textit{art} and \textit{scientist} produced by both GPT-3 and ChatGPT as shown in Figure \ref{fig:scientistart} are in stark contrast to the frame of \textit{math}, demonstrating the vast range of perceptions about academic disciplines exhibited by these LLMs. Interestingly, GPT-3 provided associations related to the \textit{mad scientist} stereotype \cite{toumey1992moral}, which were found also in the perception that high school students had of \textit{scientist} in \cite{stella2020education}.

\subsection{Comparison with GPT-4}

At the time of writing this manuscript, GPT-4 was an unreleased product made available to paying customers by OpenAI. We sampled GPT-4 a few days before it became unavailable in Italy. Here we present the results of our experiments as they relate to the in-depth results for GPT-3 and ChatGPT discussed in the above sections.

\begin{figure}
    \centering
    \includegraphics[width=11cm]{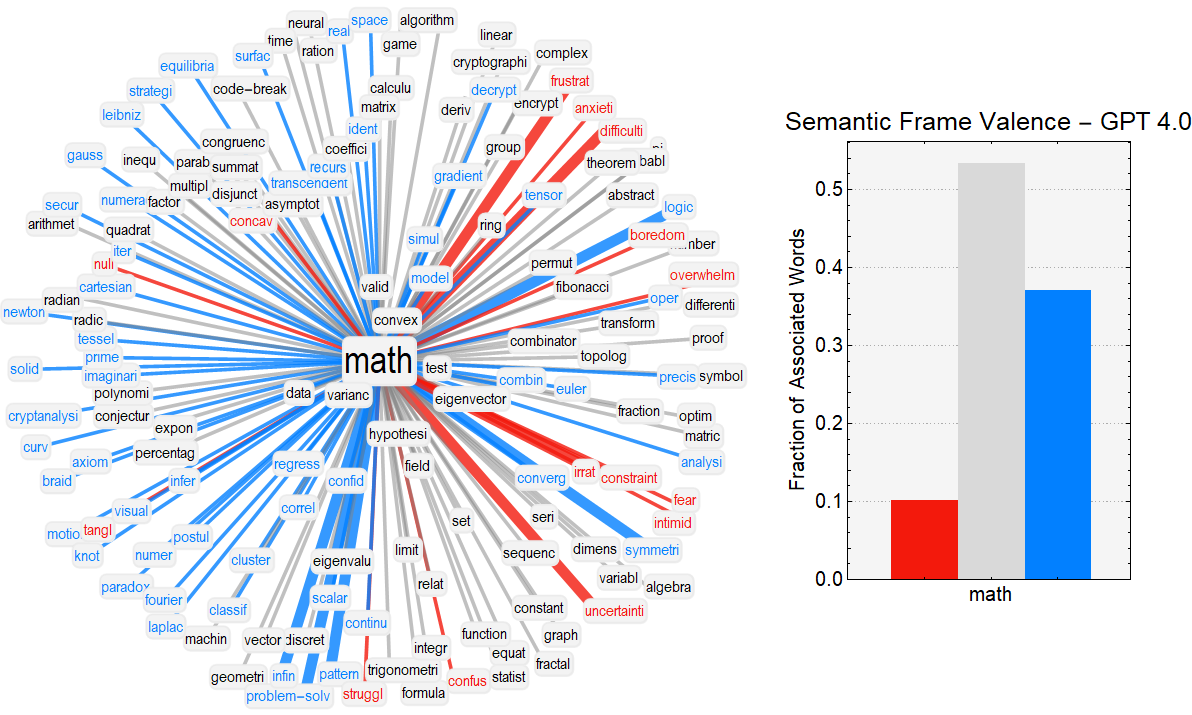}
    \includegraphics[width=11cm]{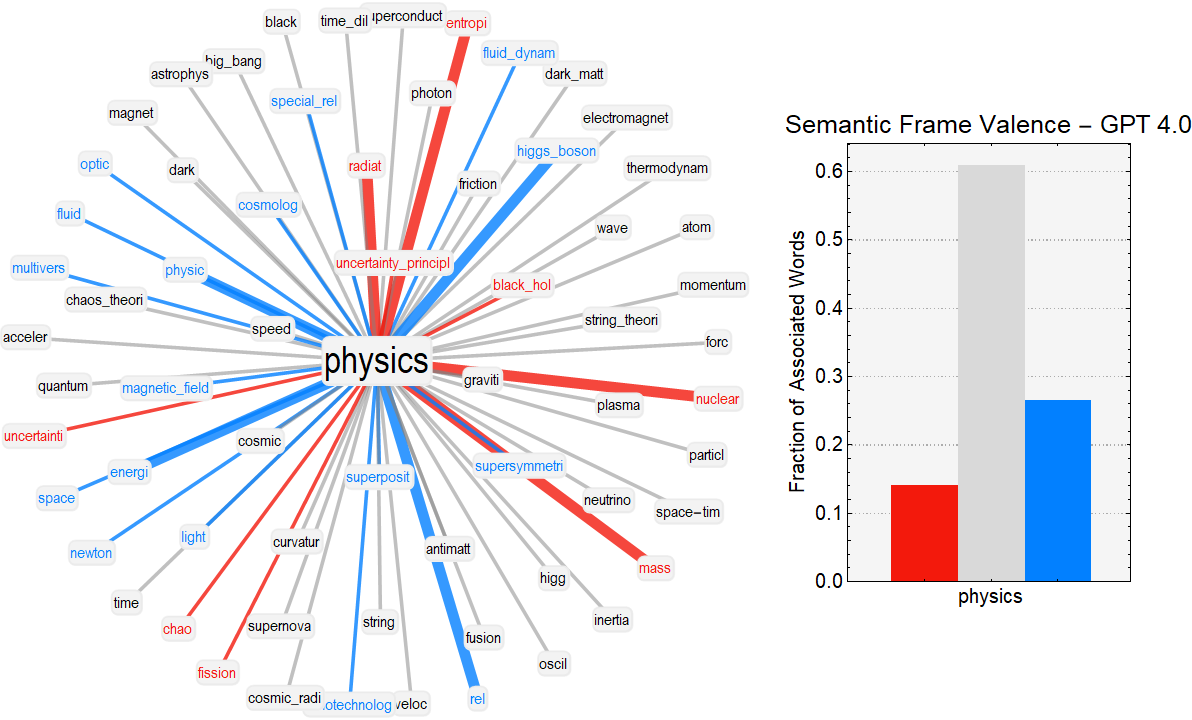}
    \includegraphics[width=11cm]{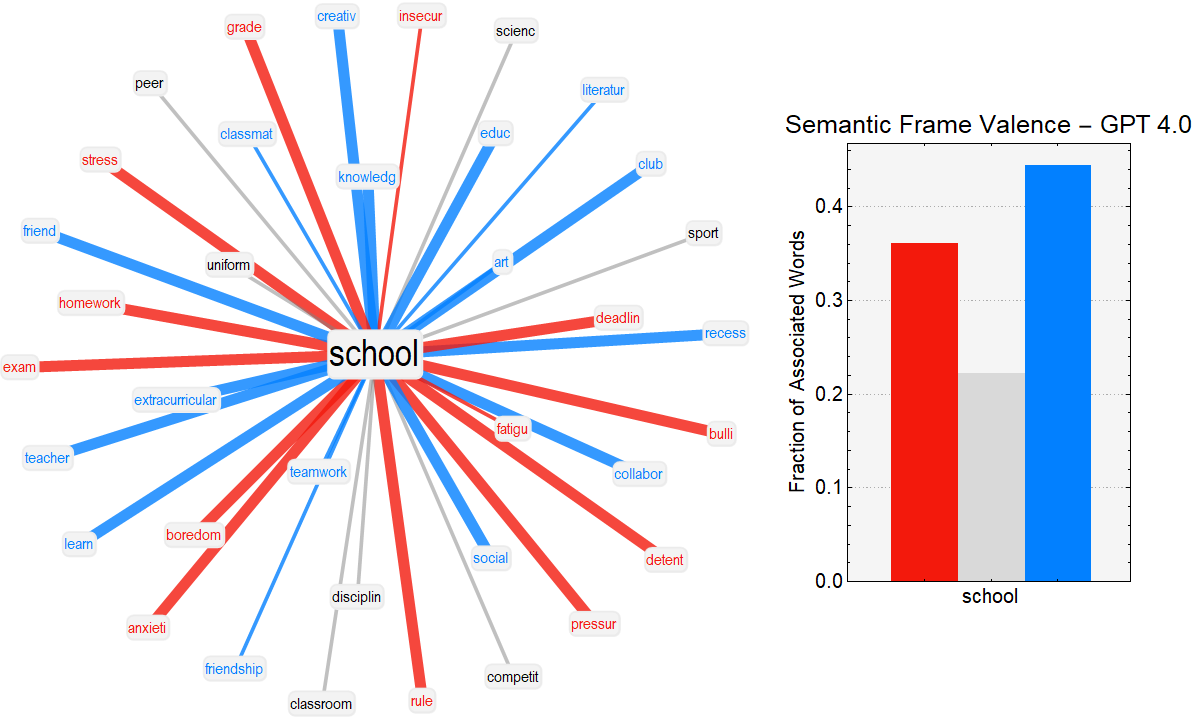}
    \caption{Semantic frames for \textit{math} (top), \textit{physics} (middle) and \textit{school} (bottom) as produced by GPT-4. The frequency of negative, positive, and neutral words in each frame are coded in frequency histograms next to each semantic frame. Color patterns are the same as in Figure 2.}
    \label{fig:gpt4}
\end{figure}

Figure \ref{fig:gpt4} reports the semantic frames of \textit{math}, \textit{physics}, and \textit{school} produced by GPT-4. Significantly notable are that these semantic frames are far more positive compared to those produced by GPT-3 and ChatGPT, especially for \textit{math} and \textit{physics}. Both \textit{math} and \textit{physics} were perceived neutrally by GPT-4 but associated mostly with positive concepts. Although negative associations constituted less than $15\%$ of the semantic frames, stereotypical associations related to math anxiety persisted. Even in the positive associations provided by GPT-4, \textit{math} was associated with \textit{frustrating}, \textit{anxiety}, \textit{fearful}, \textit{intimidating}, 
\textit{confusing} and \textit{struggle}. These negative associations were not found in the semantic frame of \textit{physics}, whose negative associates were related to domain knowledge (e.g. \textit{chaos}, \textit{nuclear}). This semantic frame analysis thus implies that negative perceptions towards \textit{math} by GPT-4 have been reduced compared to its earlier versions, but negative stereotypical perceptions persist, reflecting the psychological phenomena of math anxiety that is pervasive in society.

The semantic frame of \textit{school} produced by GPT-4 was less dominated by negative valence than the one produced by GPT-3, but was similarly polarized as the one produced by ChatGPT. Negative associations related \textit{school} with \textit{frustrating}, \textit{exam}, \textit{anxiety}, and \textit{boredom}, indicating a persistent negative perception of school settings with negative emotions and test anxiety. Interestingly, GPT-4 associated \textit{school} with \textit{bullying}, an association that was absent in results from previous language models. With GPT-4 being trained on a larger amount of web data, this association might reflect the growing sensitivity to bullying in school as discussed in online forums.

Overall, our semantic frame analysis shows that negative perceptions of math and physics exhibited by GPT-3 and ChatGPT have been reduced in GPT-4. Nonetheless, harmful stereotypical perceptions about math related to frustration and anxiety persist. This is of great concern considering the increasing use of LLMs by students. Exposure to negative implicit biases towards math in LLMs poses the risk of exacerbating the harmful math anxiety that has negative consequences on both individuals and our society.

\subsection*{A focus on the evolution of math perceptions in LLMs}

Table 1 reports the different measurements outlined in the Methods relative to the semantic frames of \textit{math} as obtained from GPT-3, ChatGPT, GPT-4, and high school students (data obtained from \cite{stella2019forma}). Interestingly, GPT models produced math-focused larger semantic frames of increasing semantic richness (i.e. number of unique associates provided for \textit{math}) in subsequent generations. This indicates a progressively richer framing/connotation of math corresponding to an increase of complexity and parameters in a LLM. Even though the sample size of responses was the same for LLMs and high school students, ChatGPT and GPT-4 produced larger semantic frames compared to high school students, i.e. more variety of responses (first row).

\begin{table}[]
\centering
\caption{Reference values for the semantic frame of \textit{math} as reported by GPT 3, ChatGPT, GPT 4, and high school students.}
\resizebox{\textwidth}{!}{%
\begin{tabular}{|l|l|l|l|l|}
\hline
\textbf{Measure/Model}            & \textbf{GPT-3} & \textbf{ChatGPT} & \textbf{GPT-4} & \textbf{High School Students} \\ \hline
Semantic Frame Size           & 30                   & 134         & 149      & 48  \\ \hline
Estimated Valence & $1.8\pm0.1$  (Negative)                  &   $2.0\pm0.1$  (Negative)        & $3.3\pm0.2$ (Neutral)   & $1.8\pm0.3$  (Negative)   \\ \hline
Estimated Frame Valence                 & Negative                  &  Negative       &   Positive &  Negative    \\ \hline
Positive/Neutral/Negative $\%$  in Frame               & 0.06/0.33/0.61                   & 0.18/0.32/0.50   & 0.37/0.53/0.10    &    0.10/0.44/0.46 \\ \hline
Non-Emotional Words in Frame              & 0.37                   &      0.56      &   0.74      & 0.84 \\ \hline
Non-Emotional W. Positive/Neutral/Negative $\%$  in Fr.             & 0.18/0.37/0.45    &         0.12/0.33/0.54      &     0.39/0.51/0.10     & 0.07/0.40/0.43  \\ \hline
\end{tabular}%
}
\end{table}

Interestingly, as summarized in Table 1 (second row), GPT-3, ChatGPT, and high school students all assigned a negative valence label to \textit{math} while GPT-4, assigned a neutral valence label. These negative and neutral perceptions are consistent with the respective negative and positive semantic frames (third/fourth row). Hence, the negative perceptions of \textit{math} by older LLMs are in line with those of high school students who are influenced by math anxiety. On the other hand, the newer language model (GPT-4) shows some improvement in this area, identifying \textit{math} as a neutral concept linked with several positively perceived concepts. This pattern indicates an intriguing evolution of GPT-4 compared to its predecessors in terms of overcoming negative attitudes toward math.

Table 1 also considers non-emotional words (fifth row), i.e. words that were featured in the semantic frame of \textit{math} but were not featured in the National Research Canada Emotion Lexicon \cite{mohammad2013crowdsourcing}. These non-emotional words are interesting because they provide a way to gauge the number of domain knowledge associates, i.e. associations of \textit{math} related to its foundational elements, instruments, and tools. Interestingly, high school students provided the highest percentage of non-emotional words ($84\%$). In LLMs, the percentage of non-emotional words increased with newer versions, demonstrated by a growing tendency for GPT-4 ($74\%$) to produce domain-knowledge associations compared to GPT-3 ($37\%$) and ChatGPT ($56\%$), thus approaching the amount of domain-knowledge produced by high school students ($84\%$).

Regarding the perceptions of these non-emotional concepts (sixth row), high school students ($43\%$), GPT-3 ($45\%$) and ChatGPT ($54\%$) all tended to perceive them negatively, indicating an unpleasant feeling about mathematical methods and instruments. In contrast, GPT-4 identified these non-emotional concepts as mostly neutral ($39\%$) and positive ($51\%$). This finding highlights GPT-4's more positive attitude towards math compared to older and simpler language models.

\section{Discussion}
\label{sec:disc}

Our findings provide compelling evidence that large language models like GPT 3, ChatGPT and even GPT-4 frame academic concepts such as math, school, and teachers with strongly negative associations. These deviations from neutrality were quantified within the quantitative framework of behavioral forma mentis networks \cite{stella2019forma,stella2020forma}, i.e. cognitive networks representing continued free association data enriched with valence scores. In the absence of impersonation, GPT-3 and ChatGPT in particular provided negative connotations for \textit{math}, perceiving it as a boring and frustrating discipline, and providing no positive associations with complex real-world applications. Unlike STEM experts, who linked creativity and real-world applications to \textit{math} (as found in previous work \cite{stella2019forma}), LLMs framed \emph{math} as detached from scientific advancements and real-world understanding. This pattern was identified in two different approaches, one leveraging semantic frame analysis \cite{stella2021mapping} and another using the circumplex model of affect \cite{posner2005circumplex}, powered through psychological data. Our analyses identified concerning deviations from neutrality in how ChatGPT and GPT-3 framed \emph{math}, highlighting negative stereotypical associations as expressed through negative emotional jargon, even in the latest GPT-4 model.

Exposure to these stereotypical associations and negative attitudes/framings could have serious repercussions. As discussed in the Introduction, LLMs act as psycho-social mirrors, reflecting the biases and attitudes embedded in the language used for training LLMs \cite{mitchell2023debate,ESOUZA2023119124,sasson2023mirror}. These models are complex enough to capture and mirror such human biases and negative attitudes in ways we do not yet fully understand \cite{anoop2022towards}. This lack of transparency translates into a relative difficulty in tracking the outcome of inquiries to LLMs: Are the framings provided by these artificial agents prevalent in the text produced by them? More importantly, could subtle and consistent exposure to such negative associations have a negative impact on some users? This represents an important research direction for future investigations of LLMs, particularly regarding the worsening of math anxiety. Social interactions with LLMs may thus exacerbate already existing stereotypes or insecurities about mathematical topics among students and even parents, analogous to the unconscious diffusion of math anxiety through parent-child interactions, as identified by recent psychological investigations \cite{soni2017role}. Negative associations of math and other concepts may be very subtle, e.g. LLMs might produce text framing math in ways that confirm students' pre-existing negative attitudes \cite{ashcraft2002math,ashcraft2005math}. They may also bolster subliminal messages that math is hard for some specific groups, influencing their academic performance through a phenomenon known in social psychology as stereotype threat (cf. \cite{stella2022network}). Such negative attitudes can have harmful effects on learning technical skills in mathematics and statistics, as evidenced by previous studies \cite{daker2021first,foley2017math} that found a negative association between math anxiety levels and learning performance in math and related courses. 

Notably, compared to ChatGPT, GPT-3 provided more negative associations and fewer positive associations for STEM disciplines like \emph{math} and \emph{physics} but also for \emph{school} and \emph{teacher}. In all these cases, the semantic frames produced by ChatGPT featured more unique associations compared to GPT-3, leading to semantically richer neighborhoods (e.g. the semantic frame of \emph{math} featured associations with several aspects of domain knowledge in ChatGPT but not in GPT-3). Hence, richer and more complex semantic representations for ChatGPT might depend on the more advanced level of sophistication achieved by its architecture, at least when compared to its predecessor GPT-3. This observation is further supported if we consider the performance provided by GPT-4, which was associated with more domain-knowledge concepts compared to previous LLMs. Noticeably, not only was the semantic frame for \emph{math} richer in GPT-4 compared to semantic frames from other LLMs, GPT-4 also overcame negative math attitudes by displaying more neutral and positive associations for that category. This makes the overall valence connotation for \emph{math} in GPT-4 much closer to the positive levels observed among STEM experts and very different from the overwhelmingly negative, displeasing attitudes observed in high school students \cite{stella2019forma}. In general, in GPT-4 the negative connotations for \emph{math}, \emph{physics}, and \emph{school} that were present in ChatGPT and GPT-3 seemed to appear drastically diminished, probably due to a combination of effects, e.g. a set of richer and more complex training resources selected by human intervention during the training phase to minimize bias, or a more sophisticated model parameterization, in which human intervention might filter our biases \cite{openai2023gpt4}. Either phenomena would consequently cause GPT-4 to have weaker manifestations of the biases encoded in previous instances of the model, i.e. GPT-4 might be mirroring different bias levels when compared to GPT-3 and ChatGPT. Intriguingly, there might also be a third phenomenon at play: the increased model complexity of GPT-4 might either make the model more "aware" of negative biases, or change the way it "relates" to math itself, leading to bias reduction in either case. Spreading awareness about math anxiety is a key first step to reducing it, mainly because acknowledging its potential psycho-social impacts could reduce the spread of negative attitudes towards math among peers, teachers, and family members \cite{stella2022network}. Recent psychological investigations of math anxiety among humans found reduced levels of math anxiety in students with stronger self-math overlap \cite{necka2015role}, i.e. a psychological construct expressing the extent to which an individual integrates math into their sense of self. Analogously to humans, GPT-4 might thus have an increased awareness of the biases related to math anxiety or a stronger self-math overlap, which would both explain the reduced levels of math-related biases observed in its semantic frames. Alas, in absence of more detailed information about the training material, filtering process, and architecture, we cannot narrow down the specific mechanisms for explaining the patterns observed here, but rather call for future research investigating these aspects in more detail.

In summary, the application of behavioral forma mentis networks to LLMs confirms the benefits of adopting a cognitive psychology approach for evaluating how large language models perceive and frame math and STEM-related concepts. In this respect, our contribution aligns with the goals of machine psychology \cite{hagendorff2023machine}, which aim to discover emergent capabilities of LLMs that cannot be detected by most traditional natural language processing benchmarks. In particular, because of the sophisticated ability of LLMs to elaborate and engage in open-domain conversations \cite{openai2023gpt4}, a structured cognitive investigation of behavioral patterns shown by LLMs appears to be natural and necessary. However, some caution should be taken when analogizing LLMs to participants in psychology experiments and then using the corresponding experimental paradigms to measure relevant emerging properties of LLMs.

Firstly, in cognitive psychology, there must be an adequate match between a given implemented task measuring a target process and the cognitive theory or model used to explain that process \cite{ashby1988toward,aitchison2012words}. For instance, past works have established a quantitative and theory-driven link between continued free association tasks -- deriving free associations between concepts -- and models based on such data whose network structure could explain aspects of conceptual recall from semantic memory \cite{de2013better,de2019small} or even higher-level phenomena like problem solving \cite{luchini2023convergent}. For instance, according to the spreading activation model established by pioneering work of psychologists Collins and Loftus \cite{collins1975spreading}, providing an individual with a cue word activates a cognitive process acting on a network representation, such that concepts are nodes linked together via conceptual associations. The activation of the node representing that given cue word facilitates a process such that activation signals start spreading iteratively through the network, diffusing or concentrating over other related nodes/concepts. Retrieval is then guided by stronger levels of activation which accumulated over other nodes (e.g. the cue \textit{book} leading to the retrieval of \textit{letter}). This spreading activation model has been extensively tested in cognitive psychology and it represents one among many potentially suitable models for interpreting free association data and their psychological nature within human beings \cite{hills2008search,siew2019spreadr}. However, in LLMs, this link between cognitive theory and experimental paradigms is mostly absent. Researchers do not yet know whether LLMs are able to approximate human semantic memory or any of its mechanisms \cite{mitchell2023debate}, mainly because LLMs are trained on massive amounts of textual data \cite{openai2023gpt4} in ways that differ greatly from the usual ways in which humans acquire language \cite{demetriou2021developmental} and its emotional/cognitive components \cite{aitchison2012words}. Furthermore, another difference is that LLMs usually combine text sources from multiple authors and can thus end up reflecting multimodal-type populations \cite{ferrara2023chatgpt}, making it extremely difficult to compare LLMs against the workings of a prototypical cognitive model at the level of an individual. In other words, there is a problematic connotation for LLMs as "artificial persona": These models can produce language in ways that appear similar to those of humans but "learn" language in a way that is much different from humans \cite{aitchison2012words}. 

Consequently, forma mentis networks in LLMs might not represent semantic frames \cite{stella2020forma,stella2020education} in ways that are analogous to how humans organize their semantic memory. This limitation strongly hampers the cognitive interpretation of semantic frames between human-generated and LLM-generated data. In fact, the main focus of this study is not to compare LLM-generated data with human-generated data, rather, the focus is on quantifying the attitudes expressed across several LLMs, and comparing how different implementations of the same overall cognitive architecture, i.e. transformer networks, represent and associate the same sets of stimuli according to the same initial prompt.

A consequence of the limited cognitive interpretation of LLM-generated data lies in the presence of an interplay between semantic and emotional aspects of memory. In humans, recent psychological studies have highlighted an interplay between retrieval processes in the categorical organization of episodic memory and the activation of related concepts in semantic memory \cite{weidemann2019neural,pulcu2017affective,de2022rethinking}. This translates into an interplay that emotions -- potentially coming from past positive, neutral, or negative episodic memories \cite{pulcu2017affective} -- might have in guiding or influencing retrieval (rather than encoding) of semantic knowledge \cite{weidemann2019neural,de2022rethinking}. Past works using behavioral forma mentis networks have shown that students and STEM experts attribute rather different affective connotations to the same concepts, particularly physics and mathematics \cite{stella2019forma,stella2020education}. Such differences could be interpreted in terms of episodic memories attributing different emotional connotations to the outputs of the recall processes activated by the continued free association task in BFMNs (see also \cite{stella2022network}). However, such an interpretation would not hold for LLMs, given their opaque structure and the uncertainty in the "cognitive" phenomena which regulate their concept retrieval \cite{mitchell2023debate}. To the best of our knowledge, no explanation of how LLMs work has yet to leverage cognitive models of human memory, mainly because of the intrinsically different ways in which humans and LLMs function. We raise this cautionary point as an encouragement for the psychology and cognitive science communities to provide novel theoretical models that go beyond the mere description of optimization processes and search in training data \cite{openai2023gpt4}, to develop frameworks that take into account the cognitive aspects of language for training data. Given that GPT-4 and its predecessors use vast amounts of human data, interpreting the cognitive structure of LLMs might lead to substantial advancements in understanding how human social cognitions are structured \cite{prates2020assessing}.

Can we ever expect future LLMs to be completely free from biases, stereotypical perceptions, and negative attitudes? Probably not. We found that GPT-4 produced fewer negative associations for \textit{math} compared to previous LLMs, so there is evidence of reduced biases. However, it is unlikely, and perhaps even undesirable, that future LLMs will be completely free from biases, at least when considering their training. According to \cite{doi:10.1080/0952813X.2023.2178517,ferrara2023chatgpt}, biases in LLMs can foster efficient algorithmic decision-making, especially when dealing with complex, unstable, and uncertain real-world environments. Furthermore, biases in the training data of LLMs can greatly boost the efficiency of learning algorithms \cite{ferrara2023chatgpt}. Unlike artificial systems, however, real people may produce biases because of three fundamental limitations of human cognition \cite{GRIFFITHS2020873}: limited time, limited computation power, and limited communication. Limited time magnifies the effect of limited computation power, and limited communication makes it harder to draw upon more computational resources, which may ultimately lead to biased behaviors. Cognitive science thus entails a kind of \emph{bias paradox}, where the two systems (artificial LLMs and human cognitive systems) apparently manifest a similar behavior (including eventual observable biases) as a result of structurally and functionally different architectures. In this way, the negative attitudes found here within LLMs should be taken with a grain of salt when compared to the negative perceptions mapped in humans in previous works \cite{stella2019forma,stella2020education,poquet2020reviewing}. Despite different psychological roots \cite{demetriou2021developmental}, the biases found here have much in common, considering the negative perceptions currently flowing online that depict math and other STEM concepts as boring, dry, and frustrating \cite{daker2021first,ashcraft2005math}. Overcoming these stereotypical perceptions will require large-scale policy decisions. Focused efforts should concentrate on reducing negative biases within LLMs, whose sphere of influence reaches an ever-increasing audience. Reducing the amount of bias present in LLMs after training is a feasible way to promote ethical interactions between humans and LLMs without perpetuating subtle negative perceptions of math and other neutral concepts.

\subsection{Conclusions}

In this work, we showed how the cognitive framework of behavioral forma mentis networks (BFMNs) can produce quantitative insights about the ways in which large language models portray specific concepts. Despite several limits to the cognitive interpretation of this approach, which is rooted in psychological theories about the nature of semantic and lexical retrieval processes in humans, BFMNs represent a powerful framework for highlighting key associations that are likely promoted by many LLMs. Here we found that different LLMs can greatly vary in the amount and type of negative, stereotypical, and biased associations they produce, indicating that machine psychology approaches like BFMNs can contribute to understanding differences in the structure of knowledge promoted across various large language models. 

\section*{Acknowledgements}

We acknowledge Clara Rastelli and Nicola De Pisapia for insightful feedback on very early stages of this project.

\bibliographystyle{plain} 
\bibliography{library} 

\end{document}